\documentclass[12pt,preprint]{aastex}
\usepackage{amsmath}
\usepackage{graphicx}
\usepackage{color}
\usepackage{url}
\usepackage{CJKutf8}
\usepackage{ulem}

\slugcomment{Accepted by \pasp}

\shorttitle{cometary nucleus extraction method}
\shortauthors{Hui \& Li}

\begin{document}

\title{
Is the Cometary Nucleus Extraction Technique Reliable?
}
\author{
\begin{CJK}{UTF8}{bsmi}
Man-To Hui (許文韜)$^{1}$ \& 
Jian-Yang Li (李荐揚)$^{2}$ 
\end{CJK}
%David Jewitt$^{1,2}$, \& 
}
\affil{$^1$Department of Earth, Planetary and Space Sciences,
UCLA, 595 Charles Young Drive East, Box 951567, 
Los Angeles, CA 90095-1567\\
}
%\affil{
%$^2$Department of Physics and Astronomy, UCLA, 
%430 Portola Plaza, Box 951547, Los Angeles, CA 90095-1547\\
%}
\affil{
$^2$Planetary Science Institute, 1700 E. Ft. Lowell Road, 
Suite 106, Tucson, AZ 85719\\
}
\email{pachacoti@ucla.edu}

\begin{abstract}

It depends. Our experiment reveals that, given an optically thin coma, generally, the smaller the signal ratio of nucleus to coma, the less reliable is the cometary nucleus-extraction technique. We strongly suggest the technique only be applied to cases where the nucleus signal occupies $\ga$10\% of the total signal wherein the bias is no more than a few percent. Otherwise there is probably no way to debias results from this technique in reality, since its reliability is highly sensitive to entangling complications, including the coma profile, and the point-spread function (PSF).

\end{abstract}


\section{\uppercase{Introduction}}

Comets are conceived to be amongst the most primitive objects in the solar system as they are generally less thermally evolved. They offer opportunities of scientific importance to help understand the formation and early history of the solar system. One of the aspects we ought to know is the size distribution of the cometary nuclei because it gives insight to their evolution. The size distribution is usually described by a simple power law as
\begin{equation}
{\rm d}n \propto R_{\rm N}^{-\mathit{\Gamma}} {\rm d}R_{\rm N}
,
\end{equation}
\noindent where $R_{\rm N}$ is the radius of a cometary nucleus, $\mathit{\Gamma}$ is the slope index, which is usually assumed to be a constant, and ${\rm d}n$ is the number of cometary nuclei having radii ranging from $R_{\rm N}$ to $R_{\rm N} + {\rm d} R_{\rm N}$. The slope index $\mathit{\Gamma}$ is particularly important, because it is associated with  evolutionary paths. For instance, Johansen et al. (2015) show that $\mathit{\Gamma} = 3.0$ will be expected if the comets formed by accretion of chondrules in the outer protoplanetary disc, in contrast to $\mathit{\Gamma} = 3.5$ for a collisional evolutionary path given the same material strength (Dohnanyi 1969). However, in cases where the collisional fragments have material strength correlated with size, the slope index varies in different size intervals (O'Brien \& Greenberg 2003). If the comets were evolved from accretion of binaries in a dynamically cold disc, the slope index is not a constant either, but changes from ${\it \Gamma} \sim 2$ ($10 \la R_{\rm N} \la 30$ km), to $\sim$5.8 ($2 \la R_{\rm N} \la 10$ km), and then to $\sim$2.5 ($0.1 \la R_{\rm N} \la 2$ km), discovered by Lamy et al. (2004) and Schlichting et al. (2013).

Revealing the size statistics for cometary nuclei is ambiguous compared to asteroids due to presence of comae. One of the methods is to observe comets at large heliocentric distances (e.g., $r_{\mathrm{H}} \gtrsim 5$ AU). However, the geometric conditions inevitably lead to faint nucleus signals, and any ongoing weak activity can easily skew estimates of nucleus sizes as well. Another way was first developed by Lamy \& Toth (1995) and improved subsequently (e.g., Lamy et al. 1998), which is termed the cometary nucleus-extraction technique. It removes contribution from the coma with some empirical models which are fitted from the observation, and measured the leftover signal from the coma-model-subtracted images. 

This technique has been widely used since it appeared in literatures (e.g., Lisse et al. 1999; Lamy et al. 2007; Fern\'{a}ndez et al. 2013; Bauer et al. 2015). In a few cases it did reveal nucleus sizes in excellent agreement with measurements in situ by spacecraft (19P/Borrelley, Lamy et al. 1998; 81P/Wild 2, Fern\'{a}ndez et al. 1999; 9P/Tempel 1, Fern\'{a}ndez et al. 2003; and 103P/Hartley 2, Lisse et al. 2009). However, in other cases it can also fail terribly (e.g., {\it Hubble Space Telescope} ({\it HST}) observations of comet C/2013 A1 (Siding Spring), J.-Y. Li, private communication; Bauer et al. 2017). Therefore, it is inevitable to question the reliability of this technique. In this paper, we endeavoured to investigate this point, and present our results.

\section{\uppercase{Method}}
\label{sec_mtd}

The basic idea of the nucleus-extraction technique is that the signal from an optically thin coma and a nucleus are separable, and that the coma profile can be fitted by some simplistic model, which can be mathematically expressed as
\begin{align}
\nonumber
F_{\rm m} \left( \rho, \theta \right) & = \left[ k_{\rm N} \delta \left(\rho \right) + k_{\rm C} \left( \theta \right) \rho^{-\gamma \left(\theta \right)} \right] \ast \mathcal{P} \\
& = k_{\rm N} \mathcal{P} + \left[ k_{\rm C} \left( \theta \right) \rho^{-\gamma \left(\theta \right)} \right] \ast \mathcal{P}
\label{eq_I}
.
\end{align}
\noindent Here $F_{\rm m}$ is the modelled flux of the comet as a function of the projected distance on the sky plane from the coma optocenter $\rho$ and the azimuthal angle $\theta$, $\delta\left( \rho \right)$ is the Dirac delta function, the symbol $\ast$ is the convolution operator, and $\mathcal{P}$ is the point-spread function (PSF) of the used optical system. In reality, shapes of PSFs can vary as a function of pixel coordinates of an image. In our experiment, PSFs remain constant across the field-of-view. We started with Gaussian PSFs.

The first term in the right-hand side of Equation (\ref{eq_I}) represents the contribution from the nucleus flux, viz., the PSF scaled by a factor $k_{\mathrm{N}}$. Signal from the coma is represented by the second term, which is assumed to be a power-law distribution in this work. The scaling factor $k_{\mathrm{C}}$ and the slope index $\gamma$ are both free parameters to be fitted from observations from a certain portion (annulus of inner and outer radii $\rho_1$ and $\rho_2$, respectively) of the coma, ideally without contamination from the nucleus signal. A crucial assumption of the method is that the portion of near-nucleus coma can be extrapolated from the coma-fitting region, regardless of what specific function is adopted to fit the coma. The coma-model image, which was constructed on a finer pixel grid with subsampling factor $\mathcal{S}$, is shifted by $\left|\Delta x \right| < 1$ and $\left| \Delta y \right| < 1$ in the subpixel grid, meaning that a total number $\left(2 \mathcal{S} + 1 \right)^2$ of coma-model images were produced. Each coma-model image was then rebinned back to the original resolution, and subsequently subtracted from the observed image. The resulting images are termed leftover images, where we measured the remaining flux presumably from the nucleus, by means of aperture photometry, whose centroid was also shifted in the subsampled pixel grid. The scaling factor for the nucleus signal $k_{\rm N}$ is then the ratio between the remaining flux and the flux of an energy-normalised PSF measured in the same photometry configuration. We then constructed scaled PSF images whose centers were determined by the subpixel coordinates of the photometry centroid, which were subsequently subtracted from the nucleus images, leaving us residual images. The goodness of fit was then calculated from summation of the residual counts weighted by flux uncertainty over the central region:  
\begin{equation}
\chi^2 \left(x_{\rm C},y_{\rm C},k_{\rm N}, x_{\rm N}, y_{\rm N} \right) = \sum_{x,y} \frac{\left( F_{\rm m} - F_{\rm o} \right)^{2}} {\sigma_{F_{\rm o}}^{2}}
\label{eq_chi2}
,
\end{equation}
\noindent where $x_{\rm C}$ and $y_{\rm C}$ are pixel coordinates of the coma peak, $x_{\rm N}$ and $y_{\rm N}$ are for the nucleus, and $\sigma_{F_{\rm o}}$ is the flux uncertainty at pixel coordinates $\left(x,y \right)$, which is computed from
\begin{equation}
\sigma_{F_{\rm o}} \left(x, y\right) = \frac{1}{t_{\rm exp}}\sqrt{\frac{1}{{\mathcal{G}}} \left[ F_{\rm o} \left(x, y\right) t_{\rm exp} + \frac{\sigma_{\mathrm{RN}}^{2}}{\mathcal{G}} \right] + \mathfrak{f}^2 F_{\rm o}^2 \left(x, y \right) t_{\rm exp}^{2}}
\label{eq_photsig}.
\end{equation}
\noindent Here, $\mathcal{G}$ and $\sigma_{\mathrm{RN}}$ are respectively the gain and readout noise of the observing CCD, $t_{\rm exp}$ is the exposure time, and $\mathfrak{f}$ represents the flat-field noise in a unit of the source signal. Our way to obtain $k_{\rm N}$ is reliable if the leftover profile is similar to that of a scaled PSF; no further aperture correction is needed. %In most cases, the occurrence of the least $\chi^2$ coincides with the most symmetric patterns in the residual images. In cases where the agreement fails, we choose the latter one as the final result. This happens rarely, only when the leftovers are morphologically different from the PSF.

\begin{figure}
\epsscale{1.0}
\begin{center}
\plotone{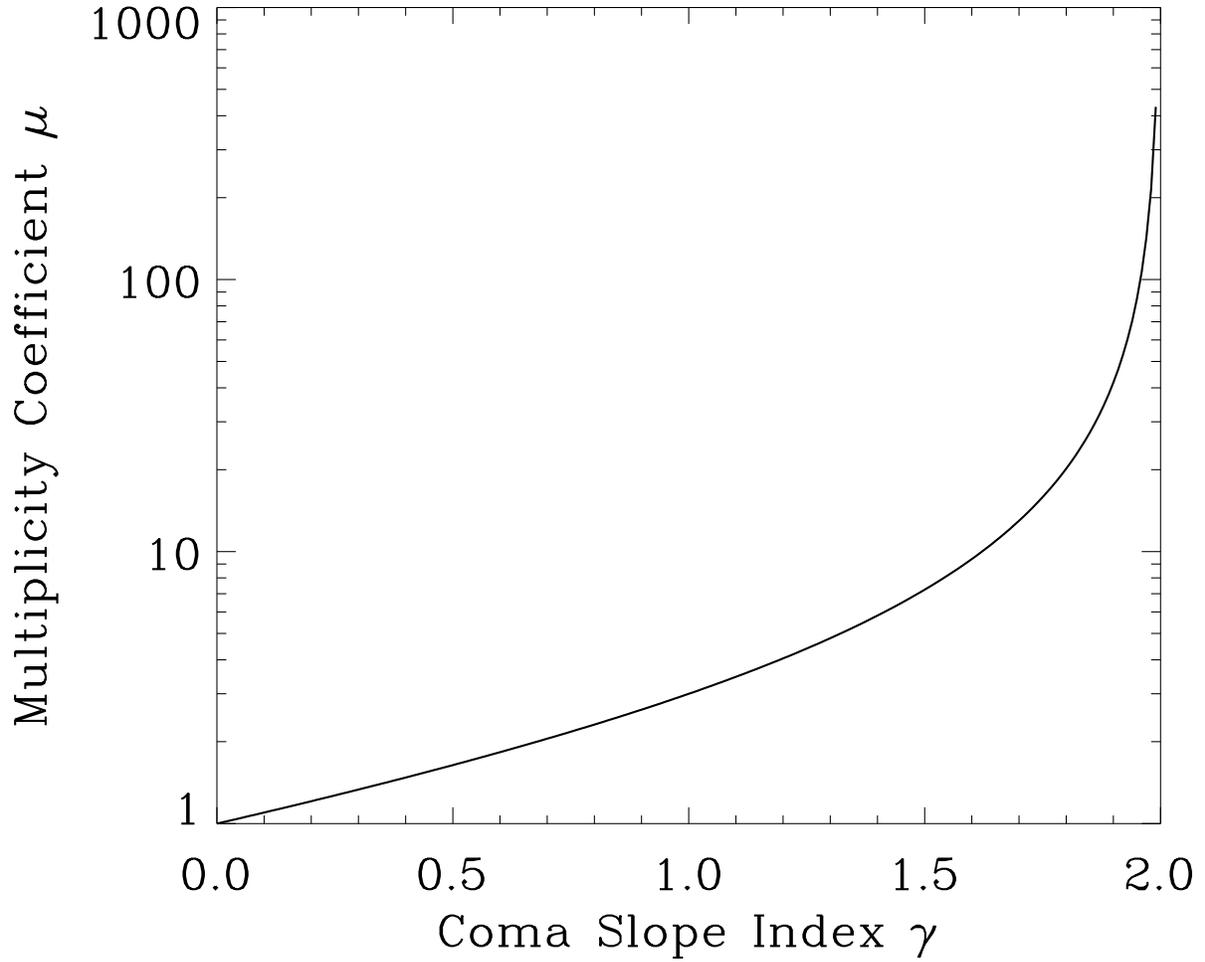}
\caption{
The multiplicity coefficient $\mu$, which is defined as the ratio between the mean pixel count at $\rho = 0$ from the peak of coma and that at the adjacent neighbouring pixel in the polar coordinates system, as a function of the coma slope index $\gamma$. According to observations of comets, typical coma slope indices remain within $1.0 \lesssim \gamma \lesssim 1.5$.
\label{fig_slope_vs_pixval}
} 
\end{center} 
\end{figure}

In this work, we opted to create a series of synthetic symmetric power-law comae with different coma-slope indices, and then added synthetic nuclei of different brightness to the synthetic images at the optocenter of comae. Noise was added to the images by arbitrarily adopting $\mathcal{G} = 1.56$ e$^{-}$/DN, $\sigma_{\rm RN} = 3.08$ e$^{-}$, $\mathfrak{f} = 0.01$, and $t_{\rm exp} = 285$ s in Equation (\ref{eq_photsig}).\footnote{These used values are typical for, e.g., {\it HST} observations (see Holtzman et al. 1995; Sirianni et al. 2005; Dressel 2012). } Our synthetic coma is circularly symmetric, i.e., no dependence upon $\theta$. A singularity exists at the optocenter of the coma ($\rho = 0$). Our solution was to compute a multiplicity coefficient $\mu$ under the polar coordinates, which is a function of the coma-slope index as
\begin{equation}
\mu \left( \gamma \right) = 2\frac{\iint_{\mathfrak{S}_0} \rho^{1-\gamma \left(\theta \right)} \mathrm{d}\theta \mathrm{d}\rho}
{\iint_{\mathfrak{S}_1} \rho^{1-\gamma \left( \theta \right)} \mathrm{d}\theta \mathrm{d}\rho }
\label{eq_sing},
\end{equation}
\noindent where region $\mathfrak{S}_0$ is defined by radii ranging from $0 \le \rho \le 1$ pixel and azimuths from $\theta$ to $\mathrm{d}\theta$, and $\mathfrak{S}_1$ has $1 \le \rho \le 2$ pixels and the same azimuthal limit as $\mathfrak{S}_{0}$. The meaning of $\mu$ is simply the ratio between the mean pixel count at the peak of the coma and that at the adjacent pixel. In cases where the coma slope index is nearly a constant, and satisfies $\gamma < 2$, Equation (\ref{eq_sing}) can be simplified to be
\begin{equation}
\mu \left( \gamma \right) = \frac{3}{2^{2 - \gamma} - 1}
\label{eq_sing_mu}.
\end{equation}
\noindent We plot $\mu$ versus $\gamma$ in Figure \ref{fig_slope_vs_pixval}. For the best-fit coma models, the replacement of the singularity was done in the same manner, yet the central pixel of the coma is assigned by a mean value:
\begin{equation}
F_{\rm C} \left( \rho = 0 \right) = \frac{\int_{0}^{2\pi} \mu \left( \gamma \right) k_{\rm C} \left(\theta \right) {\rm d} \theta}{2 \pi}
\label{eq_avg_mu},
\end{equation}
\noindent since when constructing best-fit coma models, we assumed nothing about the symmetry of the coma. Therefore the obtained $k_{\rm C}$ and $\gamma$ could vary azimuthally.

The nucleus-extraction technique was then applied on these synthetic images, whereby we obtained a set of values of the nucleus signal. We then compared the original nucleus signal and the nucleus signal extracted from the synthetic images, and evaluated the successfulness of the technique. With this approach, we quantitatively assessed how good/bad this technique is and under what conditions we can reliably extract the nucleus photometry. We arbitrarily picked $k_{\rm C} = 5$ DN s$^{-1}$, and used $\rho_1 = 10$ pixels and $\rho_2 = 90$ pixels to fit the comae (see Section \ref{sec_cfr}).

\section{\uppercase{Results}}

Our results revealed an obvious systematic bias in the nucleus-extraction technique, strongly depending on how bright the nucleus is with respect to the surrounding coma. The fainter the nucleus is, compared to the surrounding coma, the more biased the technique becomes. For this reason, hereafter we express nucleus signal in terms of parameter $\eta$, the ratio between the nucleus flux and the total flux, enclosed by a circular aperture of radius $\rho_{\mathrm{aper}} = 15$ pixels, which is arbitrarily chosen. Other factors that influence the bias include the PSF, the subsampling factor, the steepness of the coma surface brightness profile, and the coma-fitting region. In this work, the bias $\mathcal{B}$ is computed through the following equation
\begin{equation}
\mathcal{B} = \frac{k_{\rm N}^{({\rm m})} - k_{{\rm N}}^{({\rm o})}}{k_{\rm N}^{({\rm o})}} \times 100\%
\label{eq_bias},
\end{equation}
\noindent where the superscripts (m) and (o) denote the calculated and the real values, respectively. If $\mathcal{B} < 0$, the technique underestimates values for the nucleus signal.

\subsection{PSF}

\begin{figure}
\epsscale{1.0}
\begin{center}
\plotone{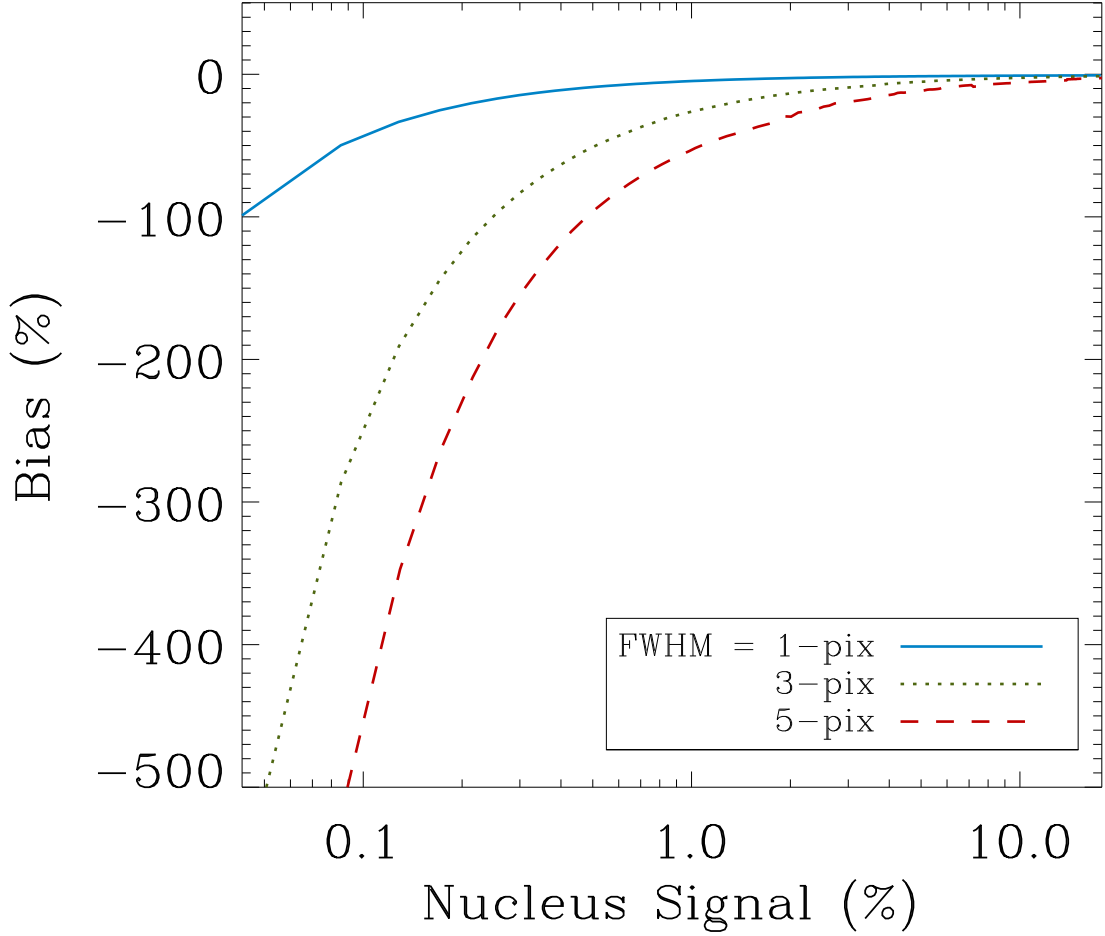}
\caption{
The systematic bias as a function of nucleus signal in terms of the ratio between the nucleus flux and the total flux enclosed by a 15-pixel radius aperture, and the FWHM of PSFs. The smaller is the FWHM of PSF, the less is the bias, which asymptotically approaches zero as the fraction of nucleus signal increases. Note that the inner radius of the coma-fitting region for FWHM = 5 pixels is changed from $\rho_1 = 10$ pixels to 20 pixels to avoid signal contamination from the synthetic nucleus. The outer radius is $\rho_2 = 90$ pixels.
\label{fig_bias_fwhm}
} 
\end{center} 
\end{figure}

We studied the effect from PSFs by choosing a narrow (${\rm FWHM} =1.0$ pixel), a moderate (3.0 pixels) and a wide (5.0 pixels) ones. For the wide-PSF case, we changed the inner radius to 20 pixels, otherwise the fitted region will be noticeably contaminated by the nucleus signal, leading to serious oversubtraction of the central region. The result is shown in Figure \ref{fig_bias_fwhm}, where we can see that for the narrow-PSF case, the systematic bias is less significant than the others. So we infer that the bias becomes worse as FWHM of PSF increases. This correlation is within our expectation, because, convolution with wider PSFs essentially blurs features more, and thus leads to loss of original information. The indication is that in reality, good seeing and sharp imaging are two of the requirements for ground-based telescopes to perform observations seeking for nucleus sizes. Without interference from the atmospheric turbulence, space telescopes such as the {\it HST} are apparently superior to the ground-based ones.

\subsection{Subsampling Factor}

\begin{figure}
\epsscale{1.0}
\begin{center}
\plotone{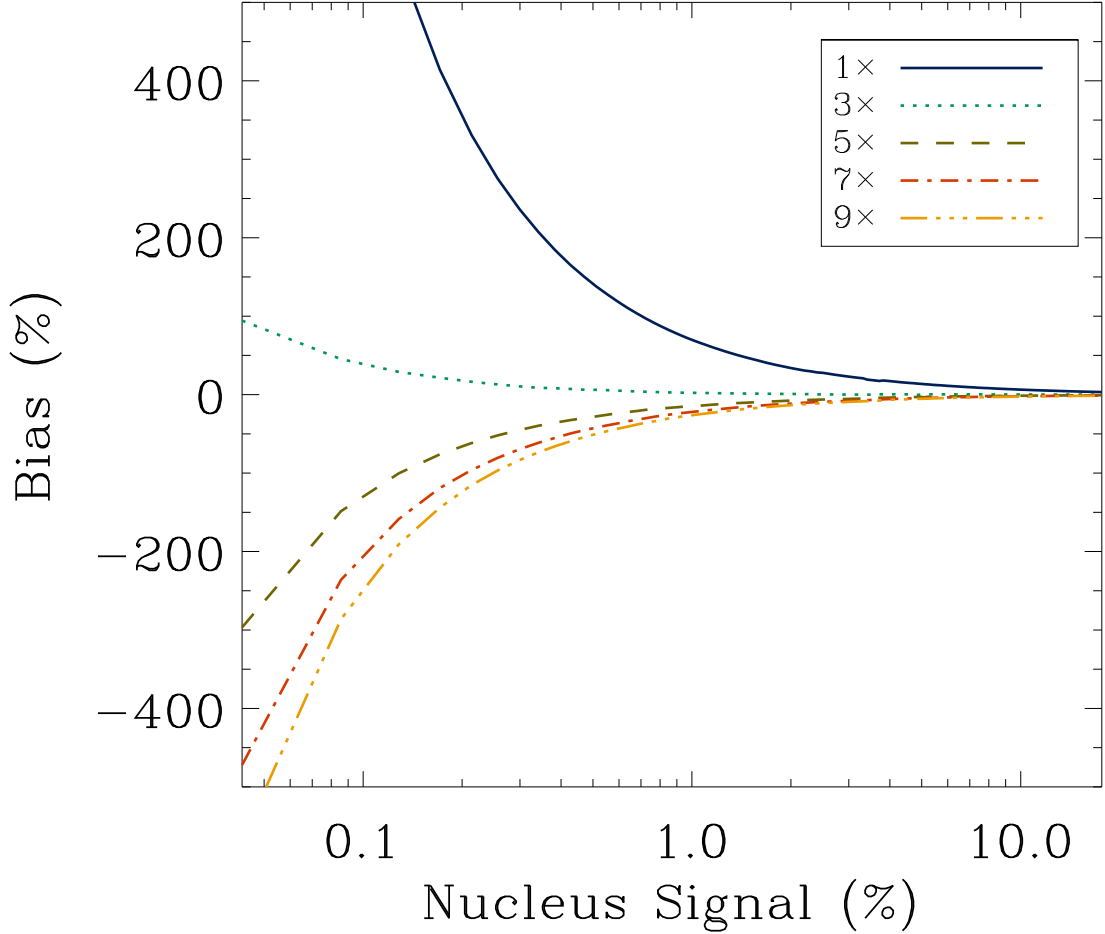}
\caption{
The systematic bias for $\rho_1 = 10$ pixels and $\rho_2 = 90$ pixels as a function of the nucleus signal and the subsampling factor for the case where the coma and nucleus are both located at the same pixel center. Note that only those with odd numbers of $\mathcal{S}$ are plotted, because the central pixel of images has the peak of coma and is the symmetric point. Nevertheless, the purpose is to demonstrate that as $\mathcal{S}$ becomes larger, the change between neighbouring bias trends shrinks. Although in this plot, the bias with $\mathcal{S} = 3$ appears to be the least, yet, once a different scheme is adopted to replace the singularity, it will be altered wildly. By contrast, the one with $\mathcal{S} = 9$ is not changed visually. Therefore, we suggest that a large subsampling factor should be used.
\label{fig_bias_rsmp}
} 
\end{center} 
\end{figure}

Researchers have been using a variety of different subsampling factors in literatures. For example, Lisse et al. (1999) used a subsampling factor of $\mathcal{S} = 5$, Lamy et al. (1998) used 8, and Li et al. (2017) used 10, etc. However, as shown in Figure \ref{fig_bias_rsmp}, in which we tested odd subsampling factors from 1 (no subsampling) to 9 with a Gaussian PSF of ${\rm FWHM} = 3$ pixels, the technique is found to have an obvious systematic bias as a function of the subsampling factor. At a first glimpse, surprisingly, the bias trend for $\mathcal{S} = 9$ is worse than that for $\mathcal{S} = 3$. We think that the origin of this problem is closely related to the arrangement of the pixel value at $\rho = 0$, which is a singularity in Equation (\ref{eq_I}). In our computation, we replaced the singularity using Equation (\ref{eq_sing_mu}). If we instead perform substitution of the singularity with the mean of the closest neighbouring pixel values, the bias trends for small $\mathcal{S}$ are impacted, but the influence dwindles as $\mathcal{S}$ increases. If the singularity is replaced by some larger number than given by Equation (\ref{eq_sing_mu}), the order of the bias trends can be completely reversed. This is to say,  the bias trend for $\mathcal{S} = 1$ becomes the worst by having the most negative values, whereas the one for $\mathcal{S} = 9$ not only becomes the best of all, but also remains largely unchanged from the one with the singularity replaced by Equation (\ref{eq_sing_mu}). Therefore, we recommend using large $\mathcal{S}$ so as to minimise effects from the way that the singularity at $\rho = 0$ is handled. However, inevitably this costs more computation time as $\mathcal{S}$ increases. 

The original purpose for subsampling is to find out the subpixel locations of coma and nucleus centers, because, in reality, they do not necessarily overlap in images, as a result from inhomogeneous activity. So, we specifically investigated this issue, by creating circularly symmetric coma models, but with non-overlapping nucleus and coma centers. We tested three different scenarios:

\begin{enumerate}

\item Nucleus shifted only, where the location of the nucleus center is shifted arbitrarily from the coma center by some subpixel displacement, whereas the center of coma remains at a pixel center.

\item Coma shifted only, where the coma center is shifted by some subpixel displacement whereas the location of the nucleus center is situated at a pixel center.

\item Both coma and nucleus shifted, by different amounts of subpixel displacements.

\end{enumerate}

\begin{figure}
\epsscale{1.0}
\begin{center}
\plotone{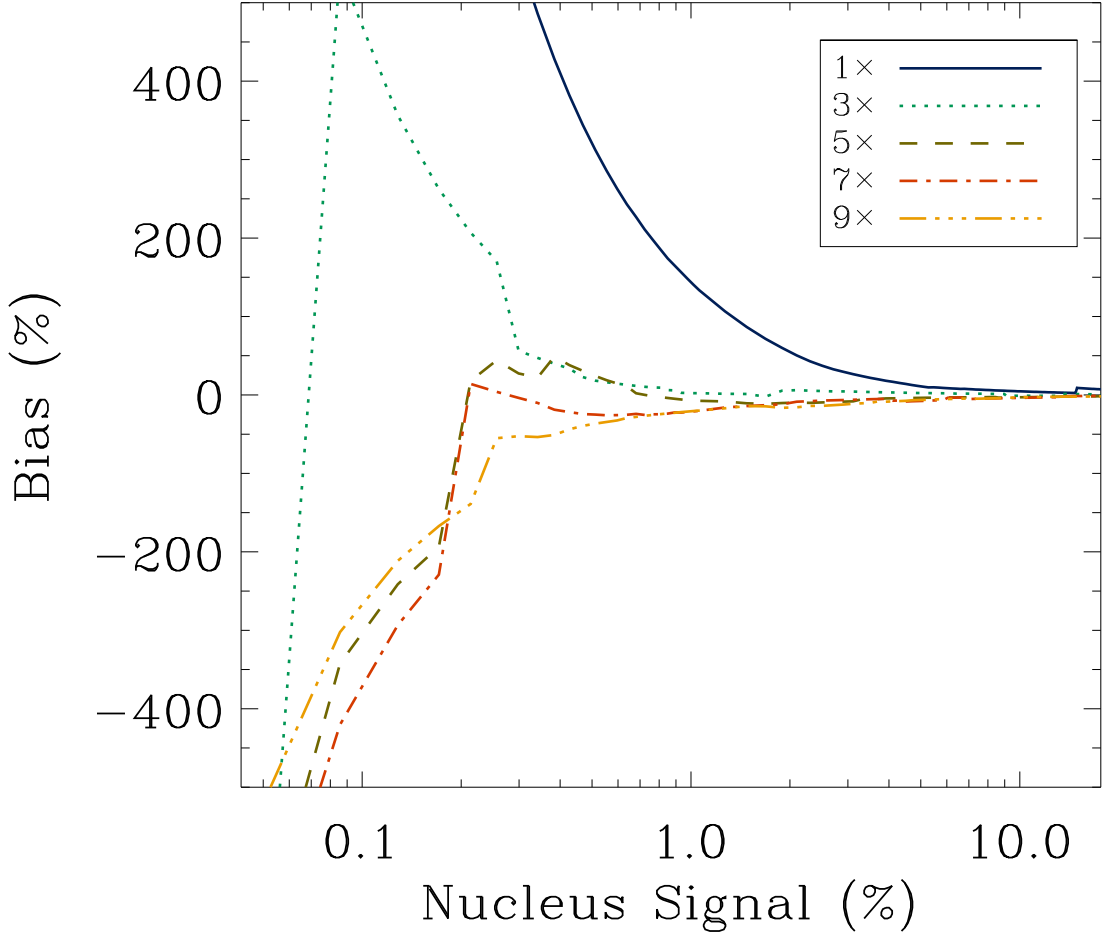}
\caption{
The same plot as Figure \ref{fig_bias_rsmp}, but with a symmetric steady-state coma with its center offset by $\left(\Delta x, \Delta y \right) = \left(-0.389, +0.452\right)$ and the nucleus shifted by $\left(+0.247, +0.113\right)$, both from a common pixel center. When the nucleus signal has $\eta \ga 1.0\%$, the shapes of the bias trends are broadly the same as in Figure \ref{fig_bias_rsmp}. However, when $\eta \la 1.0\%$, kinks due to sudden leaps in best-fit nucleus and coma centers are clearly present. Amongst the subsampling factors we tested, $\mathcal{S} = 9$ has the smallest kinks, in agreement with its leftover and residual images having the least asymmetric patterns.
\label{fig_nonsym_bias}
} 
\end{center} 
\end{figure}

We relaxed the search region for nucleus centers to $\left| \Delta x \right| \le 3$ and $\left| \Delta y \right| \le 3$, because otherwise there are cases where the obtained centers occur at the boundary of the original search region described in Section \ref{sec_mtd}. All the three scenarios were found to have broadly similar results. The bias trend for the largest subsampling factor we tested, i.e., $\mathcal{S} = 9$, is found to be the least affected, as the global shape is similar to the one in Figure \ref{fig_bias_rsmp}. The major difference that there are kinks present in Figure \ref{fig_nonsym_bias} due to the sudden jumps of the best-fitted coma and nucleus center values in the subpixel grid. The kinks are basically even more prominent for smaller values of $\mathcal{S}$, which is in line with the fact that asymmetric patterns in leftover and residual images are more severe as $\mathcal{S}$ decreases. Therefore, larger $\mathcal{S}$ ought to be adopted. 

\begin{figure}
\epsscale{1.0}
\begin{center}
\plotone{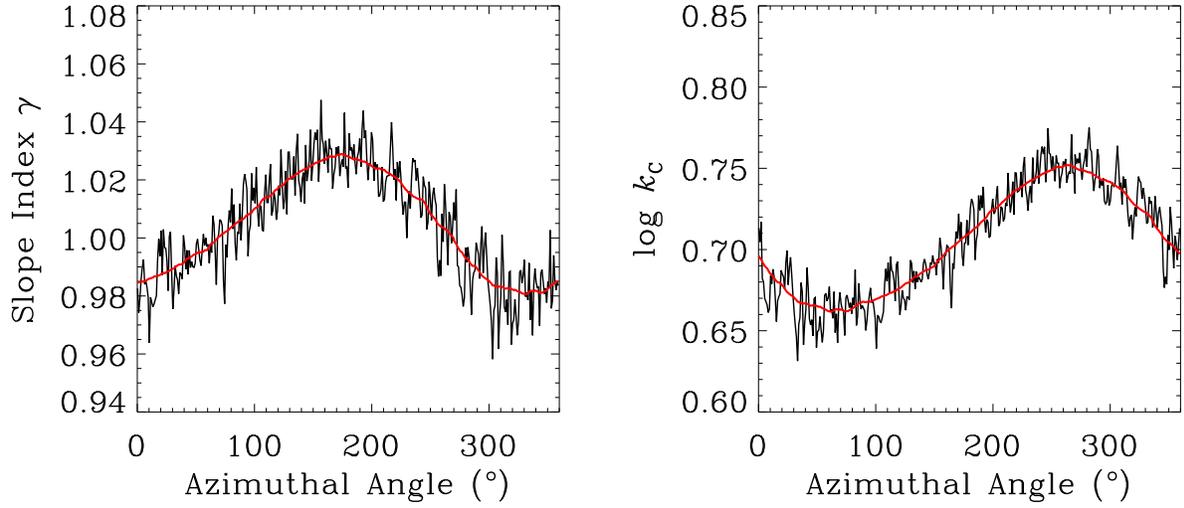}
\caption{
The best-fit parameters of the slope index $\gamma$ (left) and the scaling factor $k_{\mathrm{C}}$ (right, in logarithmic space) from an example of a circularly symmetric coma, which has a priori $\gamma = 1.0$ and $k_{\mathrm{C}} = 5.0$ DN s$^{-1}$, but has the nucleus arbitrarily shifted by $(\Delta x, \Delta y) = (+0.253, +0.617)$ from the pixel centre, which is also the coma center. The red line in each panel is the smoothed value of the corresponding parameter.
\label{fig_nonsym}
} 
\end{center} 
\end{figure}

We also found that, under no circumstances could we recover the a priori location offsets of the nucleus and coma centers from the pixel center. The reason is that, when we constructed the coma model from the best-fit parameters, the origin has already been set to the peak pixel center of the whole comet, which comprises of both the nucleus and the coma. The best-fit parameters for the coma then clearly become sinusoidal (see Figure \ref{fig_nonsym}), which are already deviated from the a priori parameters. Therefore, we conclude that, the subsampling operation fails to unravel the actual displacement information. The original purpose of subsampling cannot be fulfilled, unless for each subpixel shifted locations of the coma center, the best-fit coma model is recalculated, which is extremely time consuming and possibly unnecessary, since the bias trends for large $\mathcal{S}$ are similar to symmetric cases except for the existence of kinks. This suggests that the extraction technique is able to yield reasonably good nucleus-size estimates for comae of typical slopes, merely off by no more than a few percent, if one uses large values of $\mathcal{S}$ (e.g., $\mathcal{S} \ga 7$) when the nucleus is not too faint with respect to the surrounding coma (e.g., $\eta \ga 10\%$).

\subsection{Slope of Coma Surface Brightness}

\begin{figure}
\epsscale{1.0}
\begin{center}
\plotone{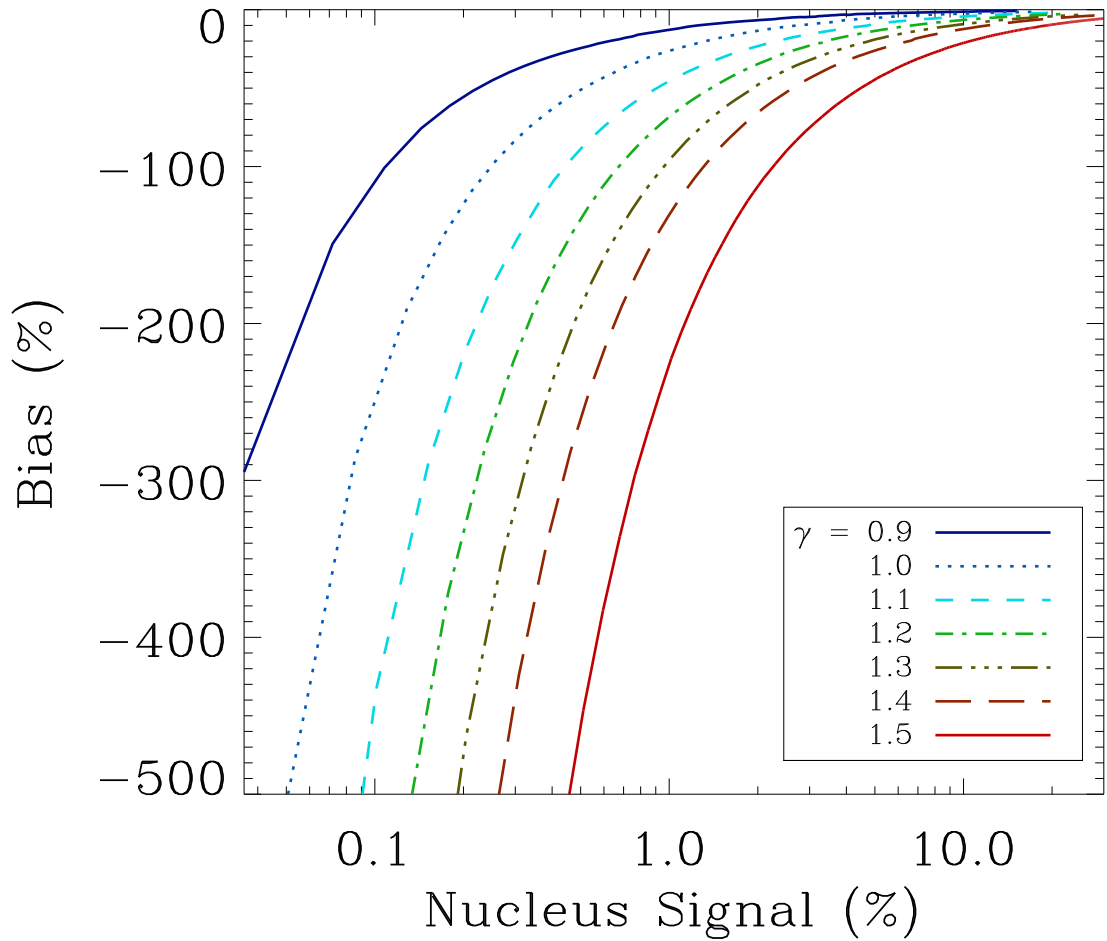}
\caption{
The systematic bias for $\rho_1 = 10$ pixels and $\rho_2 = 90$ pixels as a function of the nucleus signal and the slope index of the coma $\gamma$. As we can see, steeper slopes result in larger bias. Hypothetically, a coma with $\gamma = 0$ should have a zero bias trend regardless of the nucleus signal percentage, because the convolution operation strictly does not change the slope at all.
\label{fig_bias_slope}
} 
\end{center} 
\end{figure}

We present the systematic bias of the nucleus-extraction technique with the Gaussian PSF of ${\rm FWHM} = 3$ pixels as a function of the steepness of the coma surface brightness in Figure \ref{fig_bias_slope}. The slope index $\gamma$ varying between 0.9 (which is slightly less steep than the one in a steady-state coma, i.e., $\gamma = 1.0$) and 1.5 (corresponding to a coma under the influence of the solar radiation pressure; Jewitt \& Meech 1987), in a step of 0.1, has been tested. We fixed $\mathcal{S} = 9$. Our result is that the magnitude of systematic bias shrinks as the coma becomes less steep. The reason is that the convolution with the PSF generally has more influence upon coma surface profiles of larger values of $\gamma$. For a hypothetic coma which is completely flat across the whole image, i.e., $\gamma = 0$, convolution will not change its profile at all, so such a coma model can be accurately constructed from the observed profile without any loss, which is unfortunately not the case for comae of steeper $\gamma$. Given the coma-fitting region, as $\gamma$ increases, the technique tends to overestimate the slope index more significantly, resulting in oversubtraction of the coma.

%\section{\uppercase{Discussions}}

\subsection{Coma-Fitting Region}
\label{sec_cfr}

The bias of the nucleus-extraction technique comes into being whenever the coma profile cannot be perfectly reproduced. If the modelled coma is forced to be constructed using a priori values of parameters $k_{\rm C}$ and $\gamma$, the bias will no longer exist above the noise level. We found that how the coma-fitting region is selected strongly affects the bias trend. We thus decided to qualitatively investigate what this relationship is by varying the inner and outer radii of the annulus within which the coma profile is fitted. %We tweaked the inner radius by integer numbers of pixels, and the outer one by every ten pixels. 

We found that the trends for comae of different $\gamma$ and Gaussian PSFs of different FWHM are the same. When $\rho_1$ is small, the signal around the central region of the coma is overestimated, thereby leading to an underestimated nucleus signal. As $\rho_1$ increases, the technique then begins to systematically overestimate the nucleus signal. When the annulus is too narrow, e.g., $2\rho_1 \ga \rho_2$, the constructed coma models are no longer good approximation to the synthetic ones by being strongly asymmetric, due to the existence of noise. Before the annulus becomes too narrow, increasing $\rho_1$ whilst decreasing $\rho_2$ can reduce the systematic bias. 

\begin{figure}
\epsscale{0.9}
\begin{center}
\plotone{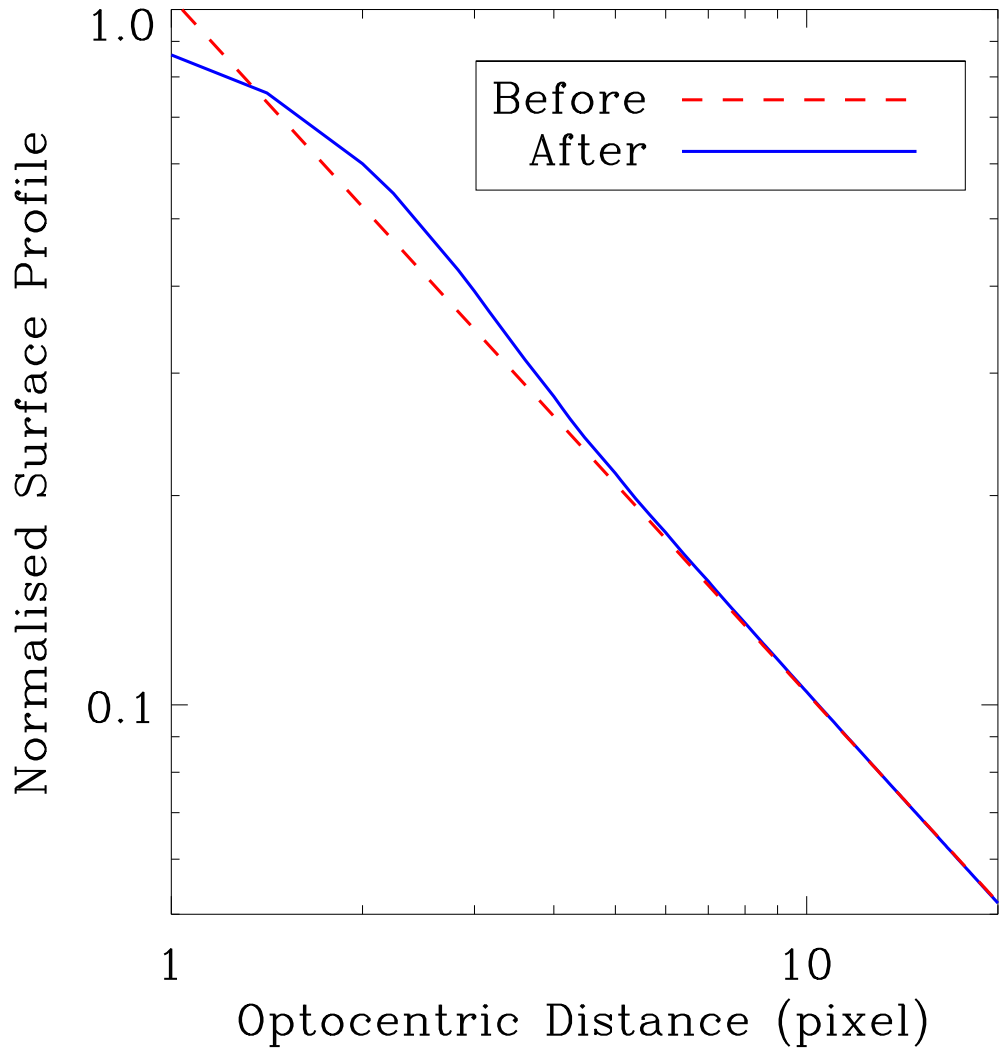}
\caption{
Comparison between the radial profiles of the $\rho^{-1}$ coma (red dashed line) before and after convolution with Gaussian PSF of FWHM $ = 3$ pixels (blue solid line), as an example. The profiles are normalised by the peak value of the post-convolution one. No nucleus is added to the model. Within $1.2 \la \rho \la 13 $ pixels, the post-convolution profile is brighter than the pre-convolution one. For $\rho \ga 13$ pixels, the post-convolution profile is always fainter than the pre-convolution one, but the difference shrinks as $\rho$ increases.
\label{fig_psfcomp}
} 
\end{center} 
\end{figure}

The behaviours can be understood from Figure \ref{fig_psfcomp}, which shows the pre-/post-convolution radial profiles of a steady state coma with the Gaussian PSF having ${\rm FWHM} = 3.0$ pixels in the logarithmic space. It is visually obvious that the slope of the post-convolution profile is steeper than the pre-convolution one when $2 \la \rho \la 10$ pixels. So if this portion of the coma is fitted, the modelled coma will then have a steeper best-fit $\gamma$, which gets even steeper after convolution with the PSF. This is the reason why smaller $\rho_1$ leads to oversubtraction of the central region. In addition to this, nucleus signal extended by PSF convolution worsens the deviation, leading to even worse oversubtraction. Starting from $\rho \sim 20$ pixels (not shown in Figure \ref{fig_psfcomp}, because the difference is extremely tiny), the slope of the post-convolution profile becomes less steep than the pre-convolution slope, and the two curves converge in an asymptotic manner. Thus, if this portion of the coma is used to construct the coma model, the central region will be underestimated. Ideally, the larger are $\rho_1$ and $\rho_2$, the better will be the best-fit coma model. However, seldom can this be realised in reality, because the signal of this portion of coma may well have insufficient SNR or can be blended with background sources. Also, comets tend to have changes in activity as functions of time, resulting in non-extrapolatable radial profiles. So in these cases the obtained coma model may be even worse than the one constructed from the portion where distortions by convolution with PSF are present.

Obviously these behaviours are highly dependent on the PSF, and also sensitive to the coma profile. So we cannot think of a simple way which can be applicable to all scenarios in reality to debias results from the nucleus-extraction technique. We did attempt to search for best-fit parameters for the synthetic comae after applying deconvolution to the synthetic comet images. However, it did not necessarily provide us with less biased results, because the noise was amplified after the operation. Neither did we determine how good the SNR should be for this procedure to work due to entangling complication factors. What we found is that for comets with typical SNR comparable to those observed by the {\it HST} (e.g., 19P, Lamy et al. 1998; C/2017 K2, Jewitt et al. 2017), deconvolution does not bring in observable improvement whatsoever, but may even deteriorate the bias. We thus conjecture that this systematic bias is probably uncorrectable. Generally speaking, we strongly recommend that, in order to obtain a reliable nucleus value, high resolution imaging about the coma with SNR as high as possible is a must. Otherwise we will expect an enormous bias stemming from the technique. %Yet in real cases where a coma is nearly free from any azimuthal variation, we conjecture that the nucleus size from the nucleus-extraction technique can possibly be debiased by running the technique on synthetic images first and then determining the corresponding systematic bias trend.

\section{\uppercase{Tests with {\it HST} Observations}}

The {\it HST} plays a unique role in measuring cometary nucleus sizes with the nucleus-extraction technique that we discussed here.  It has three advantages over almost all ground-based telescopes to apply this technique: 

\begin{enumerate}

\item The high spatial resolution attenuates the coma signal relative to the nucleus signal in the central region. 

\item The high resolution also shortens the physical distance of inner coma to be extrapolated from the coma model.

\item Resulted from being in space, it has the extremely stable PSFs, which are accurately modelled even at finer pixel grids, allowing for accurate fitting to the nucleus.

\end{enumerate}

The {\it HST} has been providing high-spatial resolution images of comets since its operation, e.g., 19P (Lamy et al. 1998), 252P (Li et al. 2017), C/2012 S1 (Lamy et al. 2014), C/2017 K2 (Jewitt et al. 2017), etc. A number of nucleus sizes or constraints have been obtained through the telescope. We thus feel the necessity to adopt the PSF of cameras WFPC2 and WFC3 onboard the {\it HST}, examine the bias trends from the nucleus-extraction technique, and also assess quality of extracted nucleus values from some of these observations.

\begin{figure}
  \centering
  \begin{tabular}[b]{@{}p{0.45\textwidth}@{}}
    \centering\includegraphics[scale=0.6]{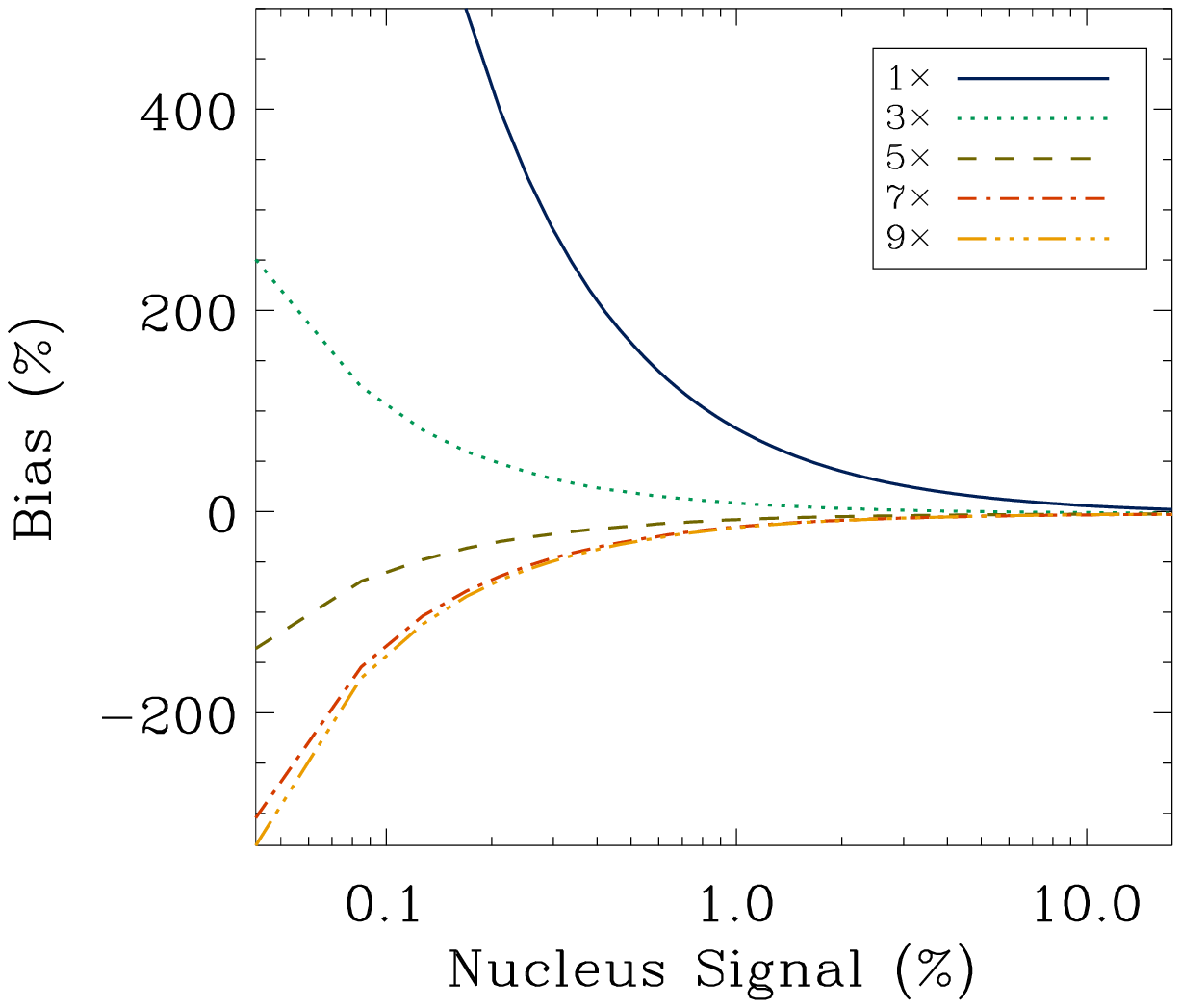} \\
    \centering\small (a)
  \end{tabular}%
  \quad
  \begin{tabular}[b]{@{}p{0.45\textwidth}@{}}
    \centering\includegraphics[scale=0.6]{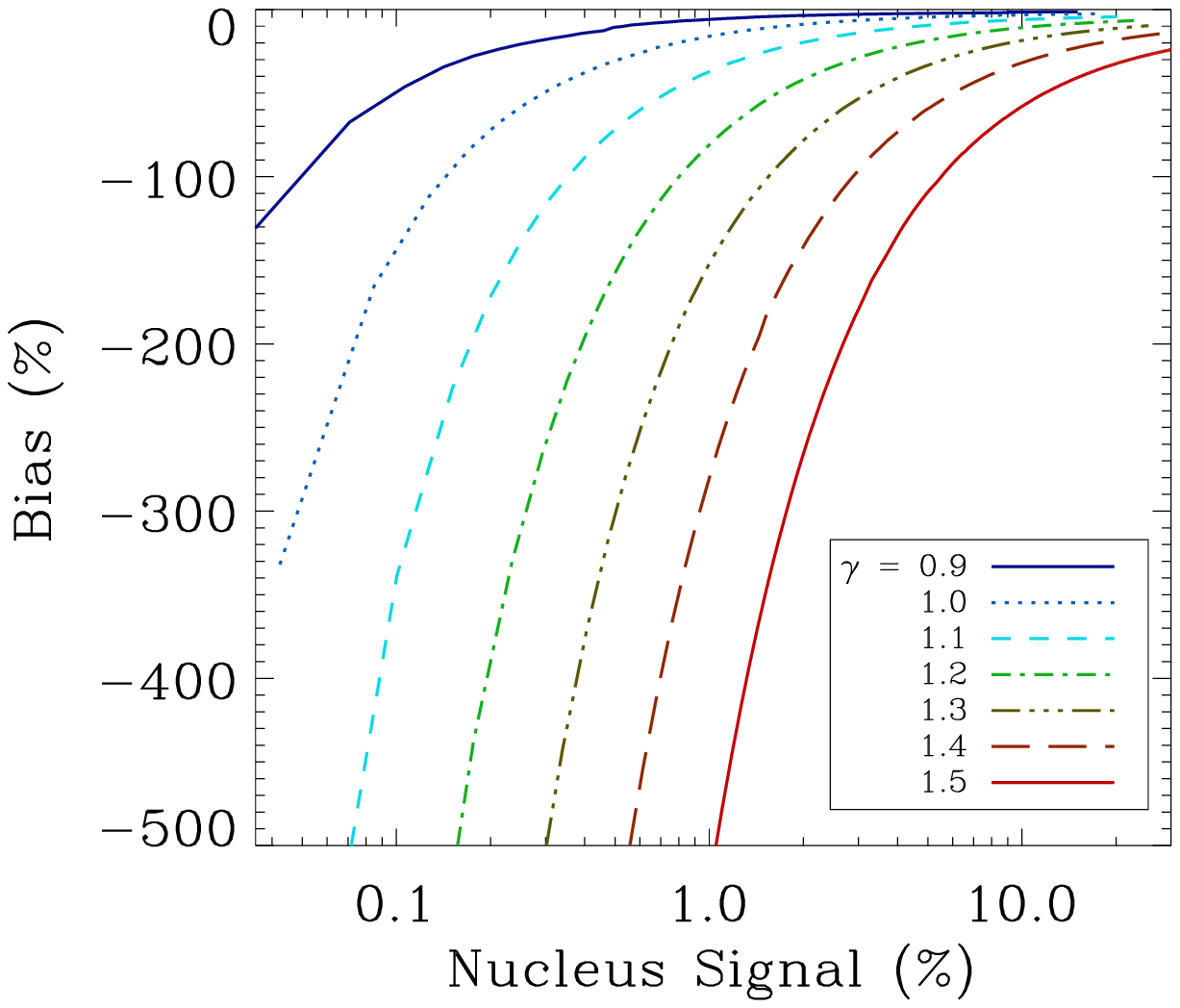} \\
    \centering\small (b)
  \end{tabular}
  \caption{
  The systematic bias trend with the {\it HST}/WFPC2 PSF as functions of the nucleus signal versus the subsampling factor (a) and the slope index of the coma (b). The coma-fitting region has $\rho_1 = 15$ pixels and $\rho_2 = 90$ pixels from the peak of the comet profile. %General trends are similar to those in Figures \ref{fig_bias_rsmp} and \ref{fig_bias_slope} with the Gaussian PSF.
  \label{fig_bias_hst_wfpc2}
  }
\end{figure}

\begin{figure}
  \centering
  \begin{tabular}[b]{@{}p{0.45\textwidth}@{}}
    \centering\includegraphics[scale=0.6]{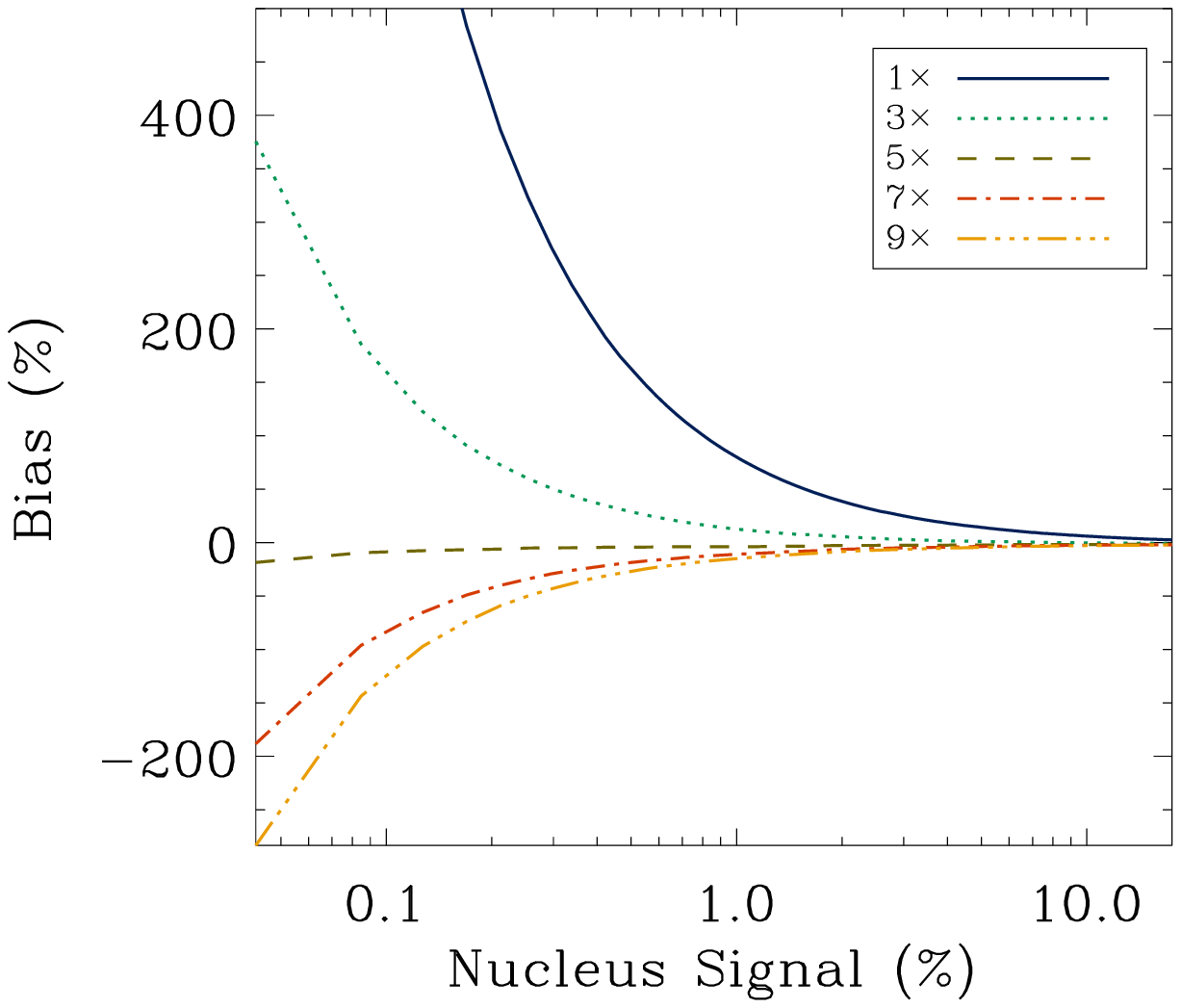} \\
    \centering\small (a)
  \end{tabular}%
  \quad
  \begin{tabular}[b]{@{}p{0.45\textwidth}@{}}
    \centering\includegraphics[scale=0.6]{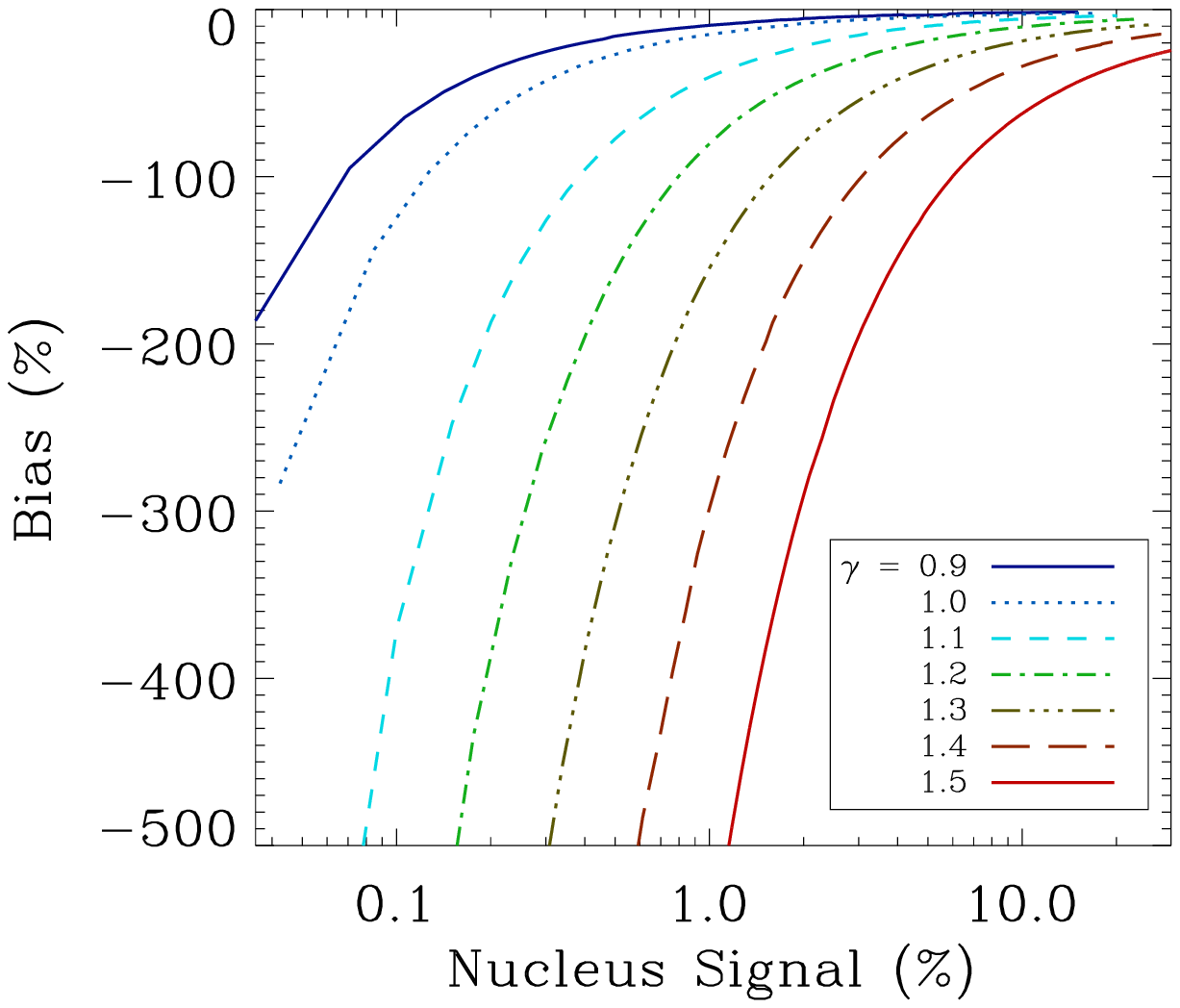} \\
    \centering\small (b)
  \end{tabular}
  \caption{
Same as Figure \ref{fig_bias_hst_wfpc2} but with the {\it HST}/WFC3 PSF. We again chose $\rho_1 = 15$ pixels and $\rho_2 = 90$ pixels from the peak of the comet profile to fit the coma. General trends are basically similar to those in Figure \ref{fig_bias_hst_wfpc2} with the {\it HST}/WFPC2 PSF.
  \label{fig_bias_hst_wfc3}
  }
\end{figure}

We performed completely the same procedures as we did for the Gaussian case on synthetic comet models with the WFPC2/WFC3 PSFs. The bias trends (see Figures \ref{fig_bias_hst_wfpc2} and \ref{fig_bias_hst_wfc3}) are generally similar to those presented in Figures \ref{fig_bias_rsmp} and \ref{fig_bias_slope}, given that a larger inner radius of the coma-fitting region ($\rho_1 = 15$ pixels for both cameras) was used. We found that the change of the bias trends with regard to the coma-fitting region differs from that in the Gaussian case. The radial profile of post-convolution image has a steeper slope starting from $\rho \la 20$ pixels, wherein it is also slightly brighter. Thus, with the used $\rho_1$ and $\rho_2$, the technique systematically oversubtracts the central region (Figures \ref{fig_bias_hst_wfpc2} and \ref{fig_bias_hst_wfc3}). Otherwise, choosing smaller radii of the coma-fitting region (e.g., $\rho_1 = 7$ pixels, $\rho_2 = 30$ pixels) leads to undersubtraction of the coma in the central region, because the slope of the post-convolution radial profile therein is shallower. Our argument that a high subsampling factor value shall be exploited mainly to avoid influence from inaccuracy of the singularity replacement is reinforced (see Figures \ref{fig_bias_hst_wfpc2}a \& \ref{fig_bias_hst_wfc3}a).

We then proceeded to assess the quality of the obtained nucleus values from several {\it HST} observations by the nucleus-extraction technique, where the nucleus sizes are known or constrained. Three comets of different activity levels are selected as representatives: 19P (weakly active), C/2013 A1 (active), and C/1995 O1 (hyperactive). Archival {\it HST} data were retrieved via the {\it HST} Moving Target Pipeline\footnote{\url{https://archive.stsci.edu/prepds/mt/}}.

\subsection{Example of Weakly Active Comet: 19P/Borrelley}
\label{sec_wac}

The size and shape of the nucleus of this comet obtained by Lamy et al. (1998) are found to be in remarkable consistence with in situ measurements by the spacecraft Deep Space 1 (see Lamy et al. 2004 and references therein). We applied the nucleus-extraction technique on one of the F675W-filtered {\it HST}/WFPC2 images from UT 1994 November 28.41. Descriptions of the observation are detailed in Lamy et al. (1998). Cosmic rays were removed by the LA Cosmic package (van Dokkum 2001) prior to applying the nucleus-extraction technique. The sky background value was computed from near-edge regions sufficiently far from the coma and then subtracted from the observed image. We varied the inner and outer radii to find best-fit parameters for the portion of coma in good SNR. The subsampling factor was fixed to be $\mathcal{S} = 9$. Table \ref{tab_19P_HST} summarizes the apparent {\it V}-band magnitudes of the nucleus of 19P with different coma-fitting region parameters, converted from the extracted fluxes by following the recipes by Holtzman et al. (1995). Estimates of the equivalent circle radius were computed by adopting an {\it R}-band geometric albedo of $0.072 \pm 0.020$, which was scaled to a {\it V}-band one with a mean color of Jupiter-family cometary nuclei ($\vr = 0.50 \pm 0.03$; Lamy \& Toth 2009; Jewitt 2015), a phase slope of 0.043 mag deg$^{-1}$, and additionally an opposition surge of 0.3 mag (Li et al. 2007). The trend of extracted values of nucleus signal versus the varying coma-fitting region is in excellent agreement with the results from our synthetic tests, i.e., less oversubtraction as $\rho_1$ and $\rho_2$ move inward to the peak of coma. By comparison, Lamy et al. (1998) obtained $R_{\rm N} = 2.12$ km for the nucleus with $\rho_1 = 7$ and $\rho_2 = 30$ pixels, which turns out to be consistent with ours (see Table \ref{tab_19P_HST}), although they exploited an optimisation approach to obtain the scaling factor $k_{\rm N}$ whilst we did not, and they adopted $\mathcal{S} = 8$, a less steep phase slope, and a lower geometric albedo. 

Nevertheless, our obtained nucleus size is consistent with the actual value ($2.17 \pm 0.03$ km, cube root of triaxial dimensions by Buratti et al. 2004). During the observation, the nucleus had $\eta \sim 10\%$ within $\rho_{\rm aper} = 15$ pixels, which marginally falls in the regime where the value obtained from the nucleus-extraction technique is less biased. In conclusion, we can see that, in cases where the comet is only weakly active, the nucleus-extraction technique is capable of rendering a reasonable estimate of the nucleus size.

\subsection{Example of Active Comet: C/2013 A1 (Siding Spring)}
\label{sec_ac}

We applied the nucleus-extraction technique on an {\it HST}/WFC3 image of the comet taken in UT 2014 March 11.11 through the F606W filter, after cosmic rays were cleaned by the LA Cosmic package. Detailed information of the observation can be seen in Li et al. (2014). Our result is that, for coma-fitting regions with large $\rho_1$ and $\rho_2$ (e.g., $\rho_1 = 10$ and $\rho_2 = 70$ pixels), we obtained an oversubtracted central region of the coma, wherein a ``blackhole" feature is present in leftover images. For small values of $\rho_1$ and $\rho_2$ (e.g., $\rho_1 = 4$ and $\rho_2 = 30$ pixels), instead a fuzzy positive leftover feature is obtained. This is completely the same as our synthetic cases where a nucleus has a tiny $\eta$ value. In order to verify this, we adopted $R_{\rm N} \sim 0.5$ km as the nucleus size of C/2013 A1 estimated by Farnham et al. (2017) from the HiRISE camera onboard {\it Mars Reconnaissance Orbiter} ({\it MRO}) during a close approach to Mars within a distance of $\sim$$1.4 \times 10^{5}$ km (Farnocchia et al. 2014). The corresponding apparent {\it V}-band magnitude of a bare nucleus of the given size having {\it V}-band geometry albedo 0.04, and phase coefficient parameter 0.04 mag deg$^{-1}$ will then be $V \approx 25.2$ during the {\it HST}/WFC3 observation. We measured the total flux centered on the peak in the {\it HST}/WFC3 image encircled by an aperture of $\rho_{\rm aper} = 15$ pixels in radius, which can be then transformed to apparent magnitude by assuming a Sun-like color. We obtained $V \approx 18.0$. Therefore, the nucleus signal in the same aperture (94.1\% of the overall) during the {\it HST}/WFC3 observation occupied merely $\sim$0.12\% of the total flux. Taking the associated uncertainties in our assumption into consideration, we remain highly confident that the nucleus signal did not exceed $\sim$0.1\%, exactly falling in the regime where the technique terribly biases actual nucleus sizes (e.g., Figure \ref{fig_bias_hst_wfc3}b). Therefore, the failure of the nucleus-extraction technique is totally within our expectation, and we envision that similar failures will occur to other active comets, unless observations are conducted during close encounters which boost the fraction of nucleus signal $\ga$10\%. Given this, we conjecture that the surprisingly flat nucleus-size distribution of long-period comets by Bauer et al. (2017) is probably not real, but due to the intrinsic bias of the technique, since, except a few (Hui 2018), the long-period comets are generally more active than short-period comets, thereby more susceptible to the bias.

\subsection{Example of Hyperactive Comet: C/1995 O1 (Hale-Bopp)}
\label{sec_hac}

\begin{figure}
\epsscale{0.8}
\begin{center}
\plotone{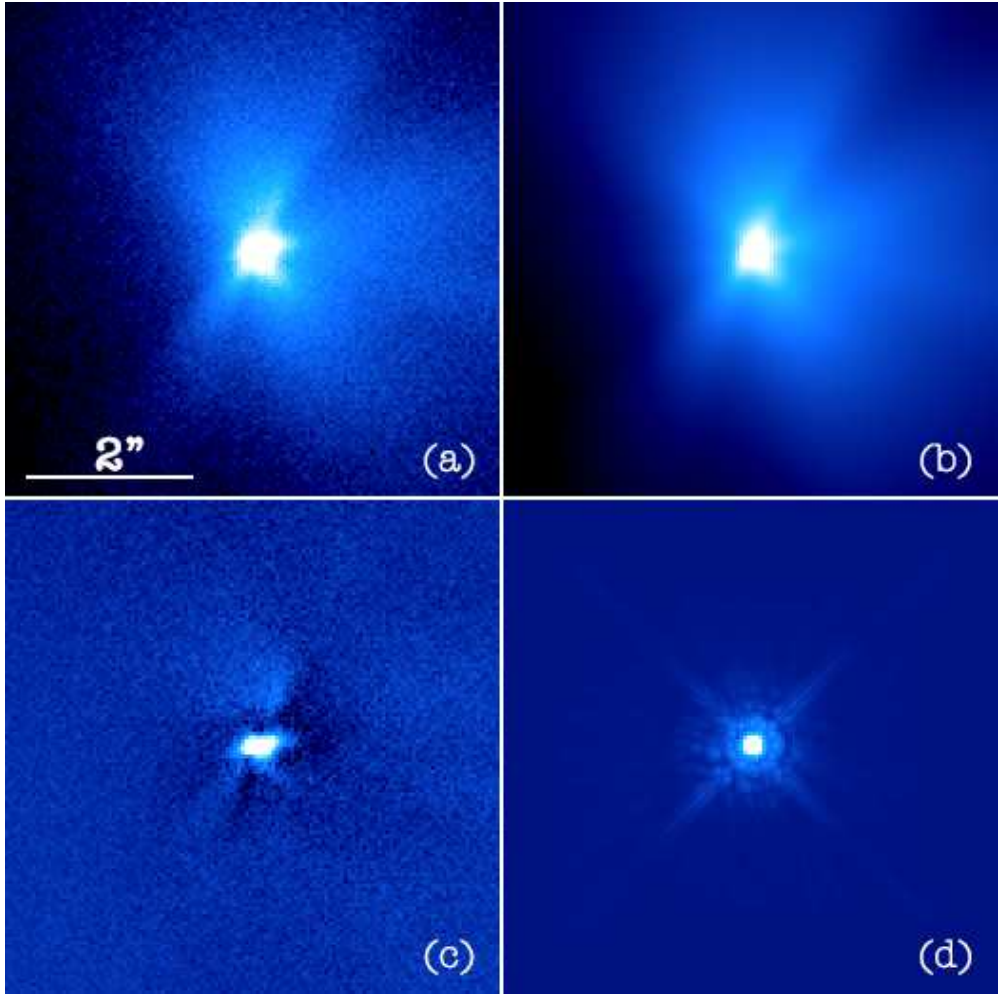}
\caption{
Comparison of the observed and model images for the {\it HST}/WFPC2 C/1995 O1 (Hale-Bopp) data from UT 1996 October 17.64. The upper left and right panels show the observed (a) and modelled (b) images, respectively, which are both stretched logarithmically in the same manner. The coma model was constructed from the annulus with $\rho_1 = 7$ and $\rho_2 = 60$ pixels. The leftover image is displayed in the lower left panel (c), whereas the lower right (d) panel is the PSF of the camera. Also shown is a scale bar. Each panel has a dimension of $5\arcsec.9 \times 5\arcsec.9$. The difference between the shapes of the leftover and PSF is readily seen, with a normalised rms of the fit $\ga$60 times more than the typical values in our experiment with synthetic comet models. 
\label{fig_HB_hst}
} 
\end{center} 
\end{figure}

We downloaded an F675W image of the comet taken by the {\it HST}/WFPC2 in UT 1996 October 17.64. Details about the {\it HST} observations are given in Weaver at al. (1997). We conducted the same procedures as in Sections \ref{sec_wac} and \ref{sec_ac}. Results of the extracted nucleus-size values from a series of inner and outer radii are listed in Table \ref{tab_HB_HST}, with the {\it V}-band geometric albedo $0.04 \pm 0.03$, the phase function coefficient $0.04$ mag deg$^{-1}$, and a mean color of nuclei of nearly isotropic comets ($\vr = 0.44 \pm 0.02$; Lamy \& Toth 2009; Jewitt 2015). Our results are consistent with previous attempts to reveal the size of the nucleus (effective radius $R_{\rm N} = 30 \pm 10$ km; Fern{\' a}ndez 2002). However, we are aware that the shapes of the leftover completely differ from the {\it HST}/WFPC2 PSF, where asymmetric patterns similar to strong cometary jets are seen in the images after the coma was subtracted (Figure \ref{fig_HB_hst}). The patterns could not be removed regardless of how we adjusted the coma-fitting region, even if an inappropriately small inner radius $\rho_1 = 4$ pixels was used. This is because the dimension of the jet features is too small compared to the annulus of the coma-fitting region (see Figure \ref{fig_HB_hst}c), thus violating the important presumption of the nucleus-extraction technique -- the near-nucleus coma shall be extrapolatable from the outer region. Therefore, solely based upon this aspect, we shall regard the nucleus-size estimates as meaningless, and the method as a failure.

Besides, the other issue that leads to the failure of the method is that comet Hale-Bopp was so active during the observation, such that its coma possibly became optically thick near the nucleus region (Weaver \& Lamy 1997). As a result, the dominant flux in the leftover is likely from strong jets in its inner coma, rather than from the nucleus, which coincides in the observed morphology of the leftover. We thus expect a much smaller nucleus size for comet Hale-Bopp than our extracted values. To conclude, this method fails to reveal the nucleus size of comet Hale-Bopp. So will it for other hyperactive comets.

\subsection{Inference on C/2017 K2 (PANSTARRS)}
\label{sec_k2}

This comet is identified as a dynamically old member from the Oort cloud, currently on its way to perihelion (Hui et al. 2017; Kr{\'o}likowska \& Dybczy{\'n}ski 2018; de la Fuente Marcos \& de la Fuente Marcos 2018). Estimates of its mass-loss rate by Jewitt et al. (2017) and Hui et al. (2017) suggest that it is activity level is ordinary in terms of a long-period comet, but on the other hand, remarkable, given the fact that it has been active when it was as far as $r_{\rm H} \approx 24$ AU, which is a record for comets in inbound legs hitherto known (Jewitt et al. 2017; Meech et al. 2017; Hui et al. 2017). Thus, we are curious about its nucleus size, and reanalysed the {\it HST}/WFC3 observations obtained by Jewitt et al. (2017) from UT 2017 June 27. The images were taken through the broadband F350LP filter and were median combined with registration on the apparent motion of the comet. We applied the nucleus-extraction technique on the coadded image using a series of $\rho_1$ and $\rho_2$. Due to the great distance of the comet ($r_{\rm H} = 15.9$ AU) during the {\it HST} observation, the angular size of the coma where the power law is still a good approximation was not big enough (angular radius $\la 2\arcsec$, or 50 pixels; Jewitt et al. 2017), so that we set $\rho_2 \le 50$ pixels as an upper boundary.

We summarized the results in Table \ref{tab_K2_HST}. The nucleus sizes were converted from the apparent magnitudes, with the assumed {\it V}-band geometric albedo 0.04 and phase slope 0.04 mag deg$^{-1}$. It seems that the effective nucleus radius of the comet is $\sim$4--5 km. However, we are aware that the obtained nucleus flux is merely $\la$0.7\% of the total flux, suggesting the unreliability of the results. If the coma is extrapolatable all the way to the near-nucleus region, the nucleus sizes revealed by the nucleus-extraction technique are expected to be underestimated. Otherwise it is unclear how the estimates are off from the actual nucleus size.

Instead, we prefer a conservative upper limit to its nucleus size using the threshold of $\eta < 10\%$. Therefore the nucleus during the {\it HST} observation was $V > 23.3$, corresponding to an equivalent circle radius of $R_{\rm N} \la 20$ km. Future high-resolution observations of the comet aiming at better constraining (or revealing) its nucleus size are certainly encouraged when the comet gets much closer to the Earth, which will potentially boost the fraction of nucleus contribution to the total signal within the same photometric aperture.

\clearpage

\section{\uppercase{Summary}}

We assessed the widely used cometary nucleus-extraction technique in a systematic way for the first time. Key conclusions are summarized as follows.

\begin{enumerate}

\item The application of the nucleus-extraction technique should be restricted to cases of optically thin comae only.

\item Nucleus signal obtained from the nucleus-extraction technique can be strongly biased. The fainter the nucleus with respect to the surrounding coma, the more biased is the extracted value. Only when the nucleus signal occupies $\ga$10\% of the total signal, can the result be trusted, as the bias will be only a few percent.

\item The bias is stemmed from distortion of the coma surface profile by convolution with PSF, which is probably uncorrectable due to noise. We recommend that the portion that is less perturbed by convolution and also has good SNR shall be used. High spatial resolution images of comets are required in order to avoid embedded biases as much as possible.

\item Large subsampling factors should be adopted to overcome inaccuracy of singularity replacement, and also to get rid of asymmetric artificial patterns otherwise present in residuals.

\item Attributed to the extremely stable and well modelled PSFs, high spatial resolution space telescopes (e.g., {\it HST}) are advantageous over ground-based ones on characterisation of non-bare cometary nuclei with the nucleus-extraction technique.

%\item The systematic bias is probably uncorrectable because it is affected and complicated by many factors, such as the shape of the PSF, the slope of the coma, and the annulus to fit the coma. %We did not succeed in finding out an iterative process that can correct the bias trend.

\end{enumerate}

\acknowledgements
{
We appreciate comments on the manuscript from Ariel Graykowski, David Jewitt and the anonymous referee. Suggestions from James Gerbs Bauer and Xinnan Du has benefited this work. M.-T.H. is financially supported by David Jewitt through a NASA grant.
}

\clearpage{}

\begin{deluxetable}{cccc}
\tabletypesize{\footnotesize}
%\rotate
\tablecaption{Results of Nucleus Extraction for 19P/Borrelley
\label{tab_19P_HST}}
\tablewidth{0pt}
\tablehead{ 
\colhead{Inner Radius $\rho_1$ (pixel)} & 
\colhead{Outer Radius $\rho_2$ (pixel)}  & 
\colhead{Apparent {\it V} Magnitude}% \\
%\colhead{(pixel)} &
%\colhead{(pixel)} &
& \colhead{Effective Radius $R_{\rm N}$ (km)}
}
\startdata

7 & 30 & $17.28 \pm 0.05$ & $1.99 \pm 0.28$ \\
 & 50 & $17.30 \pm0.05$ & $1.97 \pm 0.28$\\
 & 70 & $17.32 \pm 0.05$ & $1.95 \pm 0.28$ \\ 
 & 90 & $17.39 \pm 0.05$ & $1.89 \pm 0.27$ \\ \hline
8 & 30 & $17.30 \pm 0.05$ & $1.97 \pm 0.28$ \\
 & 50 & $17.32 \pm 0.05$ & $1.95 \pm 0.28$ \\
 & 70 & $17.35 \pm 0.05$ & $1.92 \pm 0.27$ \\
 & 90 & $17.40 \pm 0.05$ & $1.88 \pm 0.27$ \\ \hline
9 & 30 & $17.31 \pm 0.05$ & $1.96 \pm 0.28$ \\
 & 50 & $17.33 \pm 0.05$ & $1.94 \pm 0.27$ \\
 & 70 & $17.37 \pm 0.05$ & $1.91 \pm 0.27$ \\
 & 90 & $17.41 \pm 0.05$ & $1.87 \pm 0.26$ \\ \hline
10 & 30 & $17.29 \pm 0.05$ & $1.97 \pm 0.28$ \\
 & 50 & $17.34 \pm 0.05$ & $1.93 \pm 0.27$ \\
 & 70 & $17.38 \pm 0.05$ & $1.90 \pm 0.27$ \\
 & 90 & $17.43 \pm 0.05$ & $1.86 \pm 0.26$ \\ \hline

\enddata

\tablecomments{
Detailed information about the observations can be found in Lamy et al. (1998), who obtained $V = 17.38 \pm 0.04$ with $\rho_1 = 7$ and $\rho_2 = 30$ pixels. A general trend is that as the inner and outer radii grow, the flux from the extracted nucleus decreases. When the inner radius is too big, e.g., $\rho_1 = 15$ pixels, conspicuous artifact is seen in leftover images because the extrapolation of the coma profile is no longer a good approximation. The uncertainties in the apparent magnitudes are computed based on error propagation by combining values yielded by Equation (\ref{eq_photsig}), errors described in Holtzman et al. (1995), and an assumed error of its color $\sigma_{V - R} = 0.03$, which is the standard deviation of the colors of Jupiter-family cometary nuclei. Uncertainty values in $R_{\rm N}$ are propagated from the errors in the apparent magnitudes and an additional error from the geometric albedo. In comparison, the mean radius in situ measured by the Deep Space 1 is $2.17 \pm 0.03$ km (Buratti et al. 2004).
}
\end{deluxetable}

\clearpage

\begin{deluxetable}{cccc}
\tabletypesize{\footnotesize}
%\rotate
\tablecaption{Results of Nucleus Extraction for C/1995 O1 (Hale-Bopp)
\label{tab_HB_HST}}
\tablewidth{0pt}
\tablehead{ 
\colhead{Inner Radius $\rho_1$ (pixel)} & 
\colhead{Outer Radius $\rho_2$ (pixel)}  & 
\colhead{Apparent {\it V} Magnitude}% \\
%\colhead{(pixel)} &
%\colhead{(pixel)} &
& \colhead{Effective Radius $R_{\rm N}$ (km)}
}
\startdata

4 & 30 & $15.66 \pm 0.03$ & $28.6 \pm 10.7$ \\
 & 60 & $15.37 \pm0.03$ & $32.7 \pm 12.3$\\
 & 90 & $15.34 \pm 0.03$ & $33.2 \pm 12.5$ \\
 & 120 & $15.38 \pm 0.03$ & $32.6 \pm 12.2$ \\
 & 150 & $15.36 \pm 0.03$ & $32.8 \pm 12.3$ \\ \hline

7 & 30 & $15.15 \pm 0.03$ & $36.3 \pm 13.6$ \\
 & 60 & $15.01 \pm0.03$ & $38.5 \pm 14.5$\\
 & 90 & $15.07 \pm 0.03$ & $37.5 \pm 14.1$ \\ 
 & 120 & $15.10 \pm 0.03$ & $37.0 \pm 13.9$ \\
 & 150 & $15.15 \pm 0.03$ & $36.2 \pm 13.6$ \\ \hline

10 & 30 & $15.01 \pm 0.03$ & $38.7 \pm 14.5$ \\
 & 60 & $15.00 \pm0.03$ & $38.7 \pm 14.5$\\ 
 & 90 & $15.03 \pm 0.03$ & $38.3 \pm 14.4$ \\ 
 & 120 & $15.11 \pm 0.03$ & $36.9 \pm 13.8$ \\
 & 150 & $15.14 \pm 0.03$ & $36.4 \pm 13.6$ \\ \hline

\enddata

\tablecomments{
Detailed information about the observations can be found in Weaver et al. (1997). Although nucleus-size estimates are given, it is noteworthy that none of the leftovers resemble the WFPC2 PSF, as there are clear and strong spatial variations that appear to be near-nucleus jets (Figure \ref{fig_HB_hst}). Uncertainties were calculated in the same manner as described in the note of Table \ref{tab_19P_HST}.
}
\end{deluxetable}

\clearpage

\begin{deluxetable}{cccc}
\tabletypesize{\footnotesize}
%\rotate
\tablecaption{Results of Nucleus Extraction for C/2017 K2 (PANSTARRS)
\label{tab_K2_HST}}
\tablewidth{0pt}
\tablehead{ 
\colhead{Inner Radius $\rho_1$ (pixel)} & 
\colhead{Outer Radius $\rho_2$ (pixel)}  & 
\colhead{Apparent {\it V} Magnitude}% \\
%\colhead{(pixel)} &
%\colhead{(pixel)} &
& \colhead{Effective Radius $R_{\rm N}$ (km)}
}
\startdata

6 & 30 & $26.66 \pm 0.06$ & $4.2 \pm 1.1$ \\
 & 40 & $26.65 \pm0.06$ & $4.2 \pm 1.1$\\
 & 50 & $26.81 \pm 0.07$ & $3.9 \pm 1.0$ \\ \hline

7 & 30 & $26.49 \pm 0.06$ & $4.5 \pm 1.1$ \\
 & 40 & $26.49 \pm0.06$ & $4.6 \pm 1.1$\\
 & 50 & $26.78 \pm 0.07$ & $4.0 \pm 1.0$ \\ \hline

8 & 30 & $26.47 \pm 0.06$ & $4.6 \pm 1.2$ \\
 & 40 & $26.50 \pm 0.06$ & $4.5 \pm 1.1$\\
 & 50 & $26.79 \pm 0.07$ & $4.0 \pm 1.0$ \\ \hline

9 & 30 & $26.44 \pm 0.06$ & $4.7 \pm 1.2$ \\
 & 40 & $26.58 \pm 0.06$ & $4.4 \pm 1.1$\\
 & 50 & $26.80 \pm 0.07$ & $3.9 \pm 1.0$ \\ \hline

10 & 30 & $26.25 \pm 0.05$ & $5.1 \pm 1.3$ \\
 & 40 & $26.58 \pm 0.06$ & $4.4 \pm 1.1$\\
 & 50 & $26.76 \pm 0.06$ & $4.0 \pm 1.0$ \\ \hline

\enddata

\tablecomments{
Detailed information about the {\it HST} observation can be found in Jewitt et al. (2017). Although we give nucleus-size estimates, they are expected to be strongly biased, and therefore not reliable, because, if so, the nucleus only contributed a fraction of $\eta \la 0.7\%$ of the total signal. Uncertainties in the apparent magnitudes are statistical errors only, computed from Equation (\ref{eq_photsig}), which are significantly smaller than the systematic uncertainties. 
}
\end{deluxetable}

\clearpage

\end{document}